\documentclass{llncs}

\usepackage{amsmath}
\usepackage{amsfonts}
\usepackage{graphicx}
\usepackage{balance}
\usepackage[tight,footnotesize]{subfigure}
\usepackage[utf8]{inputenc}
\usepackage{url}
\usepackage{color}

\usepackage{natbib}

\makeatletter
\newif\if@restonecol
\makeatother

\usepackage[linesnumbered]{algorithm2e}
\usepackage{verbatim}
\begin{document}

\title{Hacking Smart Machines with Smarter Ones: How to Extract Meaningful Data from Machine Learning Classifiers}

\author{Giuseppe Ateniese\inst{1} \and Giovanni Felici \inst{2} \and Luigi V. Mancini\inst{1} \and Angelo Spognardi\inst{1} \and Antonio Villani\inst{3} \and Domenico Vitali\inst{1}  }
 
\institute{ 
Universit\`{a} di Roma La Sapienza, Dipartimento di Informatica
 \email{\{ateniese,mancini,spognardi,vitali\}@di.uniroma1.it}
 \and
Consiglio Nazionale delle Ricerche, Istituto di Analisi dei Sistemi ed Informatica Roma
\email{giovanni.felici@iasi.cnr.it}
 \and
 Universit\`{a} di Roma Tre, Dipartimento di Matematica
 \email{villani@mat.uniroma3.it}
 }

\maketitle
%\author[sap]{Giuseppe Ateniese}
%\ead{ateniese@di.uniroma1.it}
%\author[cnr]{Giovanni Felici}
%\ead{giovanni.felici@iasi.cnr.it}
%\author[sap]{Luigi V. Mancini}
%\ead{mancini@di.uniroma1.it}
%\author[sap]{Angelo Spognardi}
%\ead{spognardi@di.uniroma1.it}
%\author[rm3]{Antonio Villani}
%\ead{villani@mat.uniroma3.it}
%\author[sap]{Domenico Vitali}
%\ead{vialit@di.uniroma1.it}

%\address[sap]{Dipartimento di Informatica, University of Rome Sapienza}
%\address[cnr]{Consiglio Nazionale delle Ricerche, Istituto di Analisi dei Sistemi ed Informatica %Roma}
%\address[rm3]{Dipartimento di Matematica, University of Roma Tre}

\date{}
  \begin{abstract}
    Machine Learning (ML) algorithms are used to train computers to
    perform a variety of complex tasks and improve with experience.
    Computers learn how to recognize patterns, make unintended
    decisions, or react to a dynamic environment.  Certain trained
    machines may be more effective than others because they are based on
    more suitable ML algorithms or because they were trained through
    {\em superior} training sets. Although ML
    algorithms are known and publicly released, training sets may not be
    reasonably ascertainable and, indeed, may be guarded as trade secrets.
    While much research has been performed about the privacy of the elements
    of training sets, in this paper we focus our attention on ML
    classifiers and on the statistical information that can be
    unconsciously or maliciously revealed from them. We show that it is
    possible to infer unexpected but useful information from ML
    classifiers. In particular, we build a novel meta-classifier and
    train it to hack other classifiers, obtaining meaningful information
    about their training sets. This kind of information leakage can be
    exploited, for example, by a vendor to build more effective
    classifiers or to simply acquire trade secrets from a competitor's
    apparatus, potentially violating its intellectual property rights.
  \end{abstract}

\section{Introduction}
\label{sec:introduction}
Machine learning classifiers are designed to make effective and
efficient prediction of ``patterns'' from large data sets.  Many
applications have been proposed in the literature (e.g.,
%\cite{Erman-semi, MedicalApplications, MachineLearningBiology,
%  MLTrafficClassification, DTPacket, DTVoltage, svm4IDSandDT}) and
\cite{Erman-semi, MachineLearningBiology,
  MLTrafficClassification, DTPacket, DTVoltage}) and
machine learning algorithms pervade several contexts of information
technology. ML approaches (such as Support Vector machines,
Clustering, Bayesian network, Hidden Markov models, etc.)  rely on
quite distinct mathematical concepts but generally they are employed
to solve similar problems.  A machine learning algorithm consists of
two phases: \textit{training} and \textit{classification}.  During the
training, the ML algorithm is fed with a \textit{training set} of
samples. In this phase, the relationships and the correlations implied
in the training samples are gathered inside the \textit{model}.
Afterwards, the model is used during the classification phase to
classify and evaluate new data.
% widespread degli algoritmi di machine learning
ML classifiers are usually able to manage a large amount of data and to adapt to 
dynamic environments. Their versatility makes them 
suitable for several important tasks.  For example, classification and regression
models are employed to analyze current and historical trends to
make predictions in financial markets
\cite{Dhar:2011:PFM:1961189.1961191, Hiemstra:1996:LRV:229633.229643,
  Plikynas:2005:AMN:2143429.2143510}, to study biological problems
\cite{MachineLearningBiology}, to support medical diagnosis
\cite{svm4cancer, ComputerAidedDiagnosis,MedicalImaging}, to classify
network traffic or detect anomalies \cite{svm4nids,
  svm-tcp, svm4IDS, Maiolini-ssh,
  MLTrafficClassification}.
% \textbf{Controllare CITAZIONI}.

One may think that it is safe to release a classifier, whether in hardware or software, 
since intellectual property laws would prevent anyone from producing 
a similar apparatus, for example, by copying its code or its design principles.  
However, releasing a trained classifier may be subject to unexpected
information leakages that make it possible to produce 
a competitive product without violating any intellectual property rights.  \\
Let us consider, for instance, a classifier
$\mathcal{C}_a$ that is less effective than a classifier
$\mathcal{C}_b$ produced by a competitor.  The ML algorithms used in
$\mathcal{C}_b$ may be publicly available or be inferred through
reverse engineering.  For example, commercial software products for
speech recognition, such as Nuance Dragon
NaturallySpeaking~\cite{dragon},
utilize widely studied Hidden Markov Models. 
These algorithms, along with their optimizations, are well-understood and 
quite standard. Thus, the common assumption is that anyone can easily replicate
them. 
In particular, we could assume that the training set used for
$\mathcal{C}_b$ is {\em superior}, in the sense that makes
$\mathcal{C}_b$ more effective than $\mathcal{C}_a$ even though both
implement essentially the same ML algorithms. What makes
$\mathcal{C}_b$ better than $\mathcal{C}_a$ is the specific knowledge
formed during the training phase, inferred by the training set.  For
instance, a classifier that makes stock market predictions based on
neural network holds its power in the weights at its hidden layer
(see~\ref{sec:ann}). But those weights depend
exclusively on the training set, hence valuable information that
must be treasured.

Thus, it is fair to ask: Is it safe to release a profitable ML
classifier?  Would selling a software/hardware classifier reveal
concrete hints about its training set, uncovering the secrets of
its effectiveness and jeopardizing the vendor?

We show that a classifier can be hacked and that it is possible to
extract from it meaningful information about its training set.  This
can be accomplished because a typical ML classifier learns by changing
its internal structure to absorb the information contained in the
training data. In particular, we devise and train a meta-classifier
that can successfully detect and classify these changes and deduce
valuable information. However, we could not report on products
released by commercial vendors because we did not get legal
permission to hack a proprietary product. 
%{\color{green}Moreover, attacking a \emph{real} classifier would hinder the validation
%of the results due to the unavailability of the original training set.}
Nevertheless, we analyzed the same ML algorithms employed by 
commercial products. For example, we considered
the HMM-based speech recognition engine of the open-source package
VoxForge which is similar to the ones employed by commercial products, 
such as Nuance Dragon NaturallySpeking.
We note, in addition, that using open-source 
software makes our experiments easily reproducible by others.

It is important to observe that we are not interested in privacy
leaks, but rather in discovering anything that makes classifiers
better than others. In particular, we do not care about protecting the
elements of the training set. Consider the following example: a speech
recognition software recognizes spoken words better than competing
products, even though they all implement the same ML algorithms. The
training set is composed of commonly spoken words, thus it does not
make sense to talk about privacy protection.  However, we show how to
build a meta-classifier trained to reveal that, for instance, the
majority of training samples came from female voices or from voices of
people with marked accents (e.g., Indian, British, American, etc.). Then, we
can extrapolate certain hidden \textit{attributes} which are somehow
absorbed by the learning algorithm, thus possibly uncovering the
secret sauce that makes the speech recognition software stay ahead of
the competition.

Therefore the type of leakage we are interested in is quite different
than that considered in privacy preserving data mining and statistical
%databases~\cite{agrawal,chaudhuri} or differential
databases~\cite{agrawal} or differential
privacy~\cite{sulq,dworkdiff}.  Indeed, in
Section~\ref{sec:differentialprivacyhacked}, we show that a system
providing Differential Privacy is utterly insecure in our model. 

\bigskip
\noindent
{\bf Remark:} 
We introduce a novel type of information leakage and show that it is inherent to learning. 
This is far from obvious and, indeed, quite unexpected: Clearly, all learning algorithms must recognize patterns in their dataset. Thus, classifiers will inherently reveal some information. The open question is whether this information has any meaning. Indeed, classifiers are very opaque objects and make it difficult to infer anything useful at all. What we show here is that it is still possible to extract something meaningful relating to properties of the training set. This is surprising and achievable through a meta-classifier that is specially trained to expose this information. 
However, we do not attempt to formally define this new type of information leakage nor provide mechanisms to prevent it.
\subsection{Contributions}
\label{sec:motivations}
Our results evince realistic issues facing machine learning
algorithms. In particular, the main contributions of our work are:
\begin{enumerate}
\item We put forward a new type of information leakage that, to the
  best of our knowledge, has not been considered before. We show that
  it is unsafe to release trained classifiers since valuable
  information about the training set can be extracted from them.
\item We propose a way to leverage the above information leakage,
  devising a general attack strategy that can be used to hack ML
  classifiers. In particular, we define a model for a meta-classifier
  that can be trained to extract meaningful data from targeted
  classifiers.
\item We describe several attacks against existing ML classifiers: we
  successfully attacked an Internet traffic classifier implemented via
  Support Vector Machines (SVMs) and a speech recognition software
  based on Hidden Markov Models (HMMs).%%%%%% aggiungere KMEANS?? %%%%%
\end{enumerate}

We believe existing classifiers, whether commercial products or 
prototypes released to the research community, are susceptible 
to our general attack strategy. 
We put forward the importance of protecting the training set
and of the need for novel machine learning techniques that would 
prevent determined competitors from probing a ML classificator and 
learning trade secrets from it. 

\subsection{Organization of this paper}
The rest of the paper is organized as follows:
Section~\ref{sec:hack-mach-learn} describes the problem and 
introduces an attack
methodology that makes use of a ML model.
%, while Section~\ref{sec:simplemodels} details the most
%common ML models and describes how some of them can plainly reveal
%potentially valuable information about the training
%data. 
Section~\ref{sec:studies} shows how we successfully applied our
proposed methodology to hack trained SVM and HMM
classifiers.
In Section~\ref{sec:differentialprivacyhacked} we analyze the
behavior of our attack methodology when the training set is provided 
through differential privacy.
Section~\ref{sec:related} contains  some related
works. Section~\ref{sec:conclusions} concludes our work with some
remarks. %and future directions.

% 
% PROBLEM DEFINITION
% 
\section{Hacking Machine Learning classifiers}
\label{sec:hack-mach-learn}
In this paper we are interested in Machine
Learning algorithms used for classification purposes, such as  Internet
traffic classifiers, speech recognition systems, or for financial
market predictions. Our goal is to hack a trained classifier  to
obtain information that was implicitly absorbed from the
elements the classifier received as input.\\
Consider for instance the \emph{Artificial Neural Networks} (ANNs) 
based on \emph{Multi-layer perceptron} (please refer to~\ref{sec:ann} for  details about this algorithm).  
Consider a simple neural network that has to learn the identity function over a vector
of eight bits, only one of them set to 1 (this example is taken
from the popular book of Mitchell~\cite{mitchell-ml}). The network has a fixed
structure with eight input neurons, three hidden units and eight
output neurons. Using the backpropagation algorithm
over the eight possible input sequences, the network eventually learns
the target function. By examining the weights of the three hidden
units, it is possible to observe how they actually encode (in  binary)
eight distinct values, namely all  possible sequences over
three bits ($000, 001, 010,\ldots,111$). 
The exact values of the hidden units for one typical run of the 
backpropagation algorithm are shown in Table~\ref{tab:ann}.
\begin{table}
\centering

	\begin{tabular}{ c c c c c }
	Input &  & Hidden Values &  & Output\\
	\hline
	\hline
	\texttt{10000000} & $\rightarrow$ & $.89$  $.04$  $.08$ & $\rightarrow$  & \texttt{10000000} \\
	\texttt{01000000} & $\rightarrow$ & $.15$  $.99$  $.99$ & $\rightarrow$  & \texttt{01000000} \\
	\texttt{00100000} & $\rightarrow$ & $.01$  $.97$  $.27$ & $\rightarrow$  & \texttt{00100000} \\
	\texttt{00010000} & $\rightarrow$ & $.99$  $.97$  $.71$ & $\rightarrow$  & \texttt{00010000} \\
	\texttt{00001000} & $\rightarrow$ & $.03$  $.05$  $.02$ & $\rightarrow$  & \texttt{00001000} \\
	\texttt{00000100} & $\rightarrow$ & $.01$  $.11$  $.88$ & $\rightarrow$  & \texttt{00000100} \\
	\texttt{00000010} & $\rightarrow$ & $.80$  $.01$  $.98$ & $\rightarrow$  & \texttt{00000010} \\
	\texttt{00000001} & $\rightarrow$ & $.60$  $.94$  $.01$ & $\rightarrow$  & \texttt{00000001} \\
	\end{tabular}
	\caption{The weights of the hidden states, taken from Figure 4.7 of~\cite{mitchell-ml}\label{tab:ann}}
\end{table}	
Basically, the hidden units of the network were able to capture the essential information from the
eight inputs, automatically discovering a way to represent the
inputs. Thus, it is possible to extract the
(possibly sensitive) cardinality of the training set by just looking at the trained network. \\
In the following section, we describe a method to extract this type of sensitive  information. 
Namely, we show in
Section~\ref{sec:casestudysvm} that it is possible to determine if a
certain type of network traffic was included in the training
set of an Internet classifier
trained on Cisco network data flows~\cite{netflowrfc}.
Similarly, we hacked a speech recognition system and were able 
to determine the accent of speakers employed during its training.
This case study is reported in Section~\ref{sec:hmm}.

\subsection{An attack strategy}
\label{sec:attack-strategy}
In this section we devise a general attack strategy against a trained
classifier that can make an attacker able to discover some statistical
information about the training set. \\
We define the training dataset $\mathcal{D}$ as a multiset where all
the elements are couples of the form $\{ ( \vec{a},l ) | \vec{a} =
\langle a_1, a_2, \ldots, a_n \rangle \}$; to simplify, we can assume
without loss of generality that $a_i \in \{0,1\}^m$, and
$l\in\{0,1\}^\nu$. Each training element $\vec{a}$ is represented as a
vector of $n$ \emph{features} (the values $a_i$ of the vector) and has
an associated classification label $l$.  
$\mathcal{C}$ is a generic machine learning classifier trained on
$\mathcal{D}$: it could be an \emph{Artificial Neural Network} (ANN),
a \emph{Hidden Markov Model} (HMM) or a simple \emph{Decision Tree} (DT).\\
We assume that $\mathcal{C}$ is disclosed after the end of the
training phase. This means that in our model the adversary cannot
taint $\mathcal{C}$ during the learning process. Instead, we assume
that the adversary is able to arbitrarily modify the behavior of
$\mathcal{C}$ during the classification process.  In fact, when
$\mathcal{C}$ is \emph{disclosed}, it includes the set of instructions
for the classification task
% (e.g. the visiting algorithm of a decision tree for ID3) 
as well as the model definition; 
% (e.g. the decision tree itself)
hence, both the data structures and the instruction sequences are
completely in the hand of the adversary.  The assumption 
that the adversary has complete access to the
classifier is  reasonable since it
%If it is software, it is enough to look at the code (even if
%obfuscated). If it is hardware, it is possible to apply techniques of
%reverse engineering. 
is possible to extract the plain classifier also from a binary
executable through, for instance, dynamic analysis
techniques~\cite{bin-reverse}.\\
\begin{figure}
	\centering
	\includegraphics[width=0.5\textwidth]{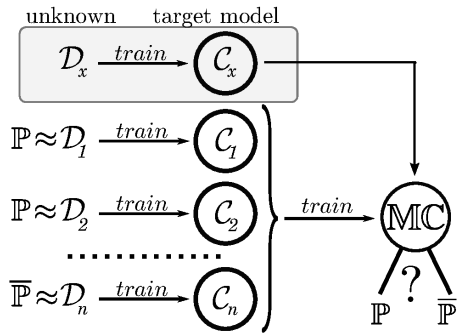}
	\caption{Attack methodology: the target training set
          $\mathcal{D}_x$ produced $\mathcal{C}_x$. Using several
          training sets $\mathcal{D}_1,\ldots,\mathcal{D}_n$ with or
          without a specific property, we build $\mathcal{C}_1,
          \ldots, \mathcal{C}_n$, namely the training set for the
          meta-classifier $\mathbb{MC}$ that will classify
          $\mathcal{C}_x$.\label{fig:sw-arch}}
\end{figure}
Each classifier $\mathcal{C}$ can be encoded in a set of feature vectors that 
can be used as input to train a meta-classifier $\mathbb{MC}$. The set of feature vectors 
that represents $\mathcal{C}$ are denoted by $\mathcal{F}_\mathcal{C}$. 
For example, in the case of an SVM, the set $\mathcal{F}_\mathcal{C}$ would contain the list of all the 
support vectors of the classifier $\mathcal{C}$.

In Figure~\ref{fig:sw-arch},  $\mathcal{C}_x$ is the trained classifier that the adversary wants
to examine in order to infer some statistical information about the training set $\mathcal{D}_x$. 
Let $\mathbb{P}$ be the property that the adversary 
wants to learn about the undisclosed $\mathcal{D}_x$. 
We write $\mathbb{P} \approx \mathcal{D}$ to say that 		
the property $\mathbb{P}$ is preserved by the dataset $\mathcal{D}$.
For instance, in the context of medical diagnosis applications, 
% the disease model, 
$\mathbb{P}$ could be: \emph{the
entries of the training set are equally balanced between males and females}.  
%To say that the property $\mathbb{P}$ is preserved by the dataset $\mathcal{D}$ we
%use the notation $\mathbb{P} \approx \mathcal{D}$.\\
To discern whether $\mathbb{P} \approx \mathcal{D}_x$, the adversary can build 
a \textit{meta-classifier} $\mathbb{MC}$, that is a classifier 
trained over a particular dataset $\mathcal{D}_\mathcal{C}$  composed of the elements $\vec{a} \in \mathcal{F}_{\mathcal{C}_i}$ labeled with $l \in \{\mathbb{P},\overline{\mathbb{P}}\}$. The label is assigned according to
the nature of the dataset used to train the classifier $\mathcal{C}_i$. \\
To train $\mathbb{MC}$ the adversary has to build the training set first. For this purpose, the adversary 
generates a vector of specific datasets 
$\vec{\mathcal{D}}=(\mathcal{D}_{1},\ldots,\mathcal{D}_{n})$ 
in such way that $\vec{\mathcal{D}}$ contains a (possibly) balanced amount of instances 
reflecting $\mathbb{P}$ and $\overline{\mathbb{P}}$.
After this step, he trains the meta-classifier $\mathbb{MC}$  as described in Algorithm~\ref{alg:train-mc}.
The algorithm takes as input the created training sets $\mathcal{\vec{D}}$ and 
their corresponding labels. 
It starts with an empty data set (line 3). Then, it trains a classifier $\mathcal{C}_i$  
on each  created data set (line 5) and gets the representation of the classifier
as a set of feature vectors (line 6). Then, it adds each feature vector to
the dataset $\mathcal{D}_\mathcal{C}$ (line 8). Finally, it trains the meta-classifier using the 
resulting data set $\mathcal{D}_\mathcal{C}$ (line 11).

\SetCommentSty{textit} 

\begin{algorithm}[t!]
  \caption{Training of the meta-classifier}
  \label{alg:train-mc}
  \small
  %\dontprintsemicolon
  
  \KwIn{\\
    $\vec{\mathcal{D}}$: the array of training sets\\
    $\vec{l}$: the array of labels, where each $l_i \in \{\mathbb{P},\overline{\mathbb{P}}\}$\\
 %   $S_m$: the specifications of $\mathcal{C}_x$\\
 %   $S_{mc}$: the specifications of $\mathbb{MC}$\\
  }
  \KwOut{The meta-classifier $\mathbb{MC}$}
  
  \BlankLine
  % \textbf{TrainMC}($\overline{\mathcal{D}}$,$\overline{l}$,$S_m$,$S_{mc}$)
  \textbf{TrainMC}($\vec{\mathcal{D}}$,$\vec{l}$)
  \BlankLine
  \Begin{
    $\mathcal{D}_\mathcal{C} = \{\emptyset\}$\\
    \ForEach{ $\mathcal{D}_{i} \in \vec{\mathcal{D}}$ } {
%      \tcp{Train the classifier with the same algorithm as
%        $\mathcal{C}_x$}
      $\mathcal{C}_i \gets$ train($\mathcal{D}_{i}$)\\
      $\mathcal{F}_{\mathcal{C}_i} \gets$ getFeatureVectors($\mathcal{C}_i$)\\
      
      % \tcp{Assign a label to each obtained feature vector and add it
      %   to the training set}
      \ForEach{$\vec{a} \in \mathcal{F}_{\mathcal{C}_i} $} {
        $\mathcal{D}_\mathcal{C}  = \mathcal{D}_\mathcal{C}  \cup \{\vec{a},l_i\}$\\ 			
      }     
    }
    \BlankLine
%    \tcp{Train the classifier with a machine learning algorithm}
    $\mathbb{MC} \gets$ train($\mathcal{D}_\mathcal{C}$)\\		
    \Return{$\mathbb{MC}$}	
  }
  \BlankLine
\end{algorithm}

Next, the adversary uses the meta-classifier $\mathbb{MC}$ on
$\mathcal{F}_{\mathcal{C}_x}$ to predict which class $l_x$ the classifier $\mathcal{C}_x$
belongs to. This is already a new form of information leakage since 
the adversary learns whether the original
training data $\mathcal{D}_x$ preserves $\mathbb{P}$ or not.\\
In practice, thanks to our attack, we are able to infer any 
key statistical property $\mathbb{P}$ preserved by the training set performing a sort of
brute-force attack on the set of properties.

It is important to remark that with this methodology the adversary
extracts \emph{external information}, NOT in the form of
attributes of the dataset $\mathcal{D}_x$. These are essentially statistical
properties inferred from the relationship among dataset entries.
For example, in Section~\ref{sec:speecstudies}
we show how to attack a speech recognition classifier by 
extracting information  about the accent of the speakers.
This information is not supposed to be captured explicitly by the model
nor it is an attribute of the training set. \\

To further improve the quality of the classification process, some
\emph{filters} can be applied to the set $\mathcal{D}_\mathcal{C}$ of
models resulting from the training phase. The filters depend on the
problem domain  and are used to find optimal models for
the property $\mathbb{P}$ and get rid of less significant entries.  In
some cases (as the example in Section~\ref{sec:casestudysvm}), this
step can be simply assimilated into the training phase of the
meta-classifier.  In other cases, as the example in
Section~\ref{sec:speecstudies}, we will discuss a filter realized with
the Kullback-Leibler
divergence~\cite{Lin91divergencemeasures}. \\
% 
% CASE STUDIES
% 
\section{Case studies}
\label{sec:studies}
In this section we provide two examples of attacks performed according
with the methodology introduced in Section~\ref{sec:attack-strategy}.
We probe two complex systems, one of which is  largely used by software vendors and
research communities.  As our first example, we attack a Speech
Recognition system realized by Hidden Markov Models; later, we
consider a network traffic classifier implemented by Support Vector
Machines. Our experiments are performed using Weka
(\cite{weka}).\\
In each experiment, we use Decision Tree as meta-classifier
$\mathbb{MC}$ (more details on Decision Tree are reported in~\ref{sec:tree}); 
we always use the $C4.5$'s implementation, namely $J48$
module, included within the Weka framework. Clearly, the attack could
be replicated using meta-classifiers based on other
ML algorithms.\\
The evaluation of our experiments is performed 
using standard metrics: (1) \textit{recall},
that is the \textit{true positive} rate, and (2) \textit{precision},
that is the ratio of true positive and the total number of positive
predictions of the model.\\
Furthermore, (3) \textit{accuracy}, namely the rate of correct
predictions made by the classifier over the number of instances of the
entire data set, can be easily derived from the confusion matrices in
Sections~\ref{sec:speecstudies} and \ref{sec:casestudysvm}.\\
In order to evaluate the effectiveness of our attack strategy, 
we crafted several classifiers trained on strongly biased training sets. These 
classifiers would probably obtain very low performance during the classification
phase; as such, they would be unlikely employed in a commercial product. 
Moreover, in our experiments, we decided to focus on simple binary properties.
Our aims are to provide an attack strategy that could be easily generalized and 
to demostrate that it is possible to infer information on the training set
looking at the weights learned by a classifier. \\
Attacking commercial products is only a matter of tuning the generation of 
the sets $\mathcal{D}_{1},\ldots,\mathcal{D}_{n}$ according to more complex 
properties. \\
To evaluate our attack strategy we make two assumptions: 1) the adversary
knows which machine learning algorithm is employed by the target 2) the adversary
has complete access to the classifier.
We claim that these two assumptions are reasonable. In fact, the information about what algorithms 
are employed is not considered a sensitive information, and sometimes it is 
advertised by the vendor itself; for instance, the newest version of the NaturallySpeaking engine 
(which is the version 12 at the time of writing) leverages HMM and five-grams 
to perform speech recognition and this information can be gathered from 
Nuance's website and patents.\\
For what concerns the second assumption, note that in many cases vendors
need to hand out their classifiers to end-users embedding them within the software executable
or apparatuses; as such, an adversary would be able
to extract the classifier using, for instance, techniques based on dynamic binary analysis.
Performing this type of analysis is orthogonal to our attack methodology and is out of 
the scope of this work.\\
It is worth remarking that the structure of the training set (e.g., the list
of attributes) is not necessary to perform our attack; indeed, we are interested on 
the \emph{external information} about the training data and we do not consider 
the attribute values.

\subsection{Hidden Markov Models}
\label{sec:hmm}
\subsubsection{Background}
A Markov Model is a stochastic process that can be represented as a
finite state machine in which the transition probability depends only
on the current state and is independent from any prior (and future)
state of the process. An \emph{Hidden Markov Model}, introduced in
\cite{baump66}, is a particular type of Markov Model for modeling
sequences that can be characterized by an underlying process
generating an observable sequence.  Indeed, only the outputs of the
states are observed (the actual sequence of the states of the process
cannot be directly observed). One of the most elegant examples to
describe HMMs was conceived by Jason Eisner \cite{eisner}:
Suppose that, in the year 2799, a climate scientist is studying the weather %AV 04/2013
in Baltimore Maryland for the summer of 2007 by examining a diary, which had 
recorded how many ice creams were eaten by Jason every day of that summer.
Only using
this record (the observable sequence), is it possible to estimate
with a good approximation the daily temperature (the hidden sequence).
HMMs solve the \emph{sequential learning problem} that is a special
learning problem where the data domain is sequential by its nature
(e.g. speech recognition problem).
\begin{figure}
	\begin{center}
	\includegraphics[width=0.5\textwidth]{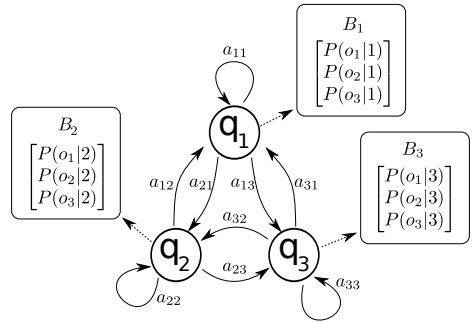}
	\caption{An example of Hidden Markov Model with three states.\label{fig:hmm-model}}
	\end{center}
\end{figure}
In Figure~\ref{fig:hmm-model}, a simple model $M$ is represented that
can be described by:
\begin{itemize}
\item a set of hidden states $Q = {q_1,q_2,...,q_m}$
\item a transition probability matrix $$A=\begin{bmatrix}a_{11} &
    a_{12} & \ldots & a_{1m}\\a_{21} & a_{22} & \ldots & a_{2m} \\
    \vdots & \vdots & \ddots & \vdots\end{bmatrix}$$ where the element
  $a_{i,j}$ represents the probability of moving from state $i$
  to state $j$%, that is  \[ a_{i,j} = P(x_t=q_j|x_{t-1}=q_i) \]
\item an emission probability matrix $B(m\times n)$, where the element
  $b_{j,k}$ is the probability to produce the observable $o_k$ from
  the state $j$, that is
  \[
  b_{j,k}=B_j(k)=P(o_k |q_j)
  \]
\end{itemize}
The HMM model is based on two main assumptions. The first is the
Markov assumption, namely that given a sequence $x_1,\ldots, x_{i-1}$
of transitions between states, the probability of the next state
depends only on the present state:
\[
P(x_i=q_j|x_1,x_2,\ldots,x_{i-1})=P(x_i=q_j|x_{i-1})
\]
The second is the output independence assumption, namely that given a
sequence $x_1,\ldots, x_{T}$ of transitions between states, where
$x_i=q_j$, and the observed sequence $y_1,\ldots, y_{T}$, the emission
probability of any observable $o_k$ depends only on the present state
and not on any other state or observable:
\[
P(y_{i}=o_k|x_1,\ldots,x_{i},\ldots, x_T,y_1,\ldots, y_T)=P(o_k|q_j)
\]
In Figure~\ref{fig:hmm-model}, three states ($q_1$, $q_2$ and $q_3$) are shown:
the transition probabilities $a_{ij}$, and, for the three states, the
emission probabilities ($B_1,B_2,B_3$ respectively) of the three
observable ($o_1,o_2,o_3$).

The HMM models are well-suited to solve three types of problems:
likelihood, decoding and learning \cite{Jurafsky}.  Likelihood
problems are related to evaluating the probability of observing a given
observable sequence $y_1,\ldots,y_T$, given a complete HMM model,
where both matrices $A$ and $B$ are known.
% (in the above example, to
%compute the probability of given a sequence of ice creams eaten by
%Jason). 
Decoding problems call for the evaluation of the best sequence of hidden states
$x_1,\ldots,x_T$ that can have produced a given observable sequence
$y_1,\ldots,y_T$. 
%(for example to compute the most probable daily
%temperature given a sequence of ice creams ate by Jason). 
Learning
problems consist of reconstructing the two matrices $A$ and $B$ of an HMM,
given the set of states $Q$ and one (or more) observation sequence
$Y$.
For this task, the \emph{Viterbi} and the \emph{Baum-Welch} algorithms are used respectively
to train and tune the HMM. 

\medskip
\subsubsection{HMM for speech recognition} 
\label{sec:speecstudies}
In this section we describe the attack to the HMM in the specific case of 
\emph {Speech Recognition Engines} (SRE).
\emph{Speech Recognition} (SR) is the process of converting a
sound recorded through an acquisition hardware to a sequence of written
words. The applications of SR are manyfold: dictation, voice
search, hands-free command execution, audio archive searching, etc.
The predominant technology used to perform this task is the HMM~\cite{hmmspeech},
many tools are nowadays available (\cite{julius,sphinx}).\\
We exploited our methodology to verify whether the HMM was 
trained with a \emph{biased} training set: according to the methodology
described in \ref{sec:attack-strategy}, we are able to detect with high 
confidence whether the HMM was trained \emph{only} with people from 
the same nationality. 
To recognize a speech, SREs require two types of input:
\begin{itemize}
\item an \emph{Acoustic Model}, which is created by taking speech audio
	files, i.e., the \emph{speech corpus}, and their transcriptions, and combing them
	into a statistical representation of the sounds that make up each
	word;
\item and either a \emph{Language Model} or a \emph{Grammar File}. Both describe
  the set of words that the statistical model will be able to classify.
  However, the first model contains the probabilities of sequences
  of words, while the second contains a set of predefined combinations
  of words. In the following experiment, this paper uses only the Language
  Model.
\end{itemize}
Let us briefly introduce the typical SRE workflow.
An unknown speech waveform is captured by the acquisition hardware, the
Pulse Code Modulation 
provides the digital representation of the analogical audio signal.
This bitstream is now converted in \emph{mel-frequency cepstral coefficients} 
(MFCCs), namely a representation of the short-term power spectrum of sounds. 
The MFCCs are the \emph{observables} of a Hidden Markov Model that
changes state over time and that generates one (or more) \emph{observables} 
once it enters into a new state.\\
In this scenario, the states of the HMM are all the possible subphonemes of 
the language while the transition matrix contains the probability for each 
subphoneme to cycle over itself or to move to the next subphoneme. 
The emission probabilities are the probability to observe a certain MFCC 
from each subphoneme.
The only possible transitions between the states of each phonemes are to 
themselves or to successive states, in a left-to-right fashion; the 
self-loops makes it possible to deal with the variable length of each phoneme with ease. 
Both transition and emission probabilities are built using the \emph{Viterbi}
algorithm~\cite{viterbi} over a large speech corpus.\\
Since the MFCC files are vectors of real-valued numbers, they are approximated
by the multivariate Gaussians distribution (note that the
probability to have exactly the same vector would be nearly 0). 
For any different state (i.e., subphoneme), each dimension of the vector has a certain 
\textit{mean} and \textit{variance} that represent the likelihood of an individual 
acoustic observation from that state.\\
For the sake of our experiments, we build the Hidden Markov Models
using the \emph{Hidden Markov Model Toolkit} (HTK)~\cite{htk} toolkit. 
HTK consists of a set of library modules and tools available in C. 
The HTK toolkit provides a high level of modularity and is organized through 
a set of libraries with functions (e.g., \emph{HMem} for 
memory management, \emph{HSigP} for signal processing ,\ldots) and a 
small core.
The MFCC files were gathered from the VoxForge project~\cite{wwwvoxforge}, 
the most important speech corpus and acoustic model repository for open-source
speech recognition engines.
Moreover, each speech file released by VoxForge is associated with several 
categories such as gender, age range, and pronunciation dialect.
The aim of our experiment is to extract this information, which is implicitly
correlated with the contents, even if it does not appear as an attribute in our
data set.

\subsubsection{Attack description}
\label{par:sr-attack}
The main objective of this attack is to build a meta-classifier for
the following property $\mathbb{P}$: \emph{the classifier was trained 
only with people who speak an Indian english dialect}. 
We emphasize that this is  \emph{external information} as introduced
in Section \ref{sec:attack-strategy}: the speech dialect is NOT explicitly used 
during the training process, but in practice it influences the output of the classifier.\\
The first part of the experiment describes the encoding of the
HMMs; next, we describe the decision tree of the meta-classifier; finally, 
we present an improved version of the classifier that uses a \emph{filter} 
to improve the classification.\\
To carry out the attack, we retrieved $11,137$ recordings from the VoxForge corpus. In
particular, for our experiment, we took only the MFCC files in
the English language. Each track comes with a form containing some meta-information
(e.g. gender, age, pronunciation dialect). We have partitioned the corpus according
to this meta-information; for this experiment, we have considered the partition 
containing the recordings made with the same pronunciation dialect and similar
recording equiments. 
We preprocessed the corpus with the HTK toolkit in order to minimize 
the environmental noise.
Starting from this partition, we have created $\mathcal{\vec{D}}$ according to the
rule defined in Section~\ref{sec:attack-strategy}. Then, we have trained each classifier 
$\mathcal{C}_i$ as described in Algorithm~\ref{alg:train-mc}.\\
After that,  we started with the encoding phase which is described below.
Each classifier $C_i$, is represented in the HTK toolkit by an ASCII file
containing an HMM for each phoneme belonging to the English
language. Each HMM is composed of: a transition probability matrix
$A(n \times n)$ which describes the transition between hidden states
and the two vectors $M = (\mu_1,\mu_2, \ldots , \mu_m)$ and $V =
(\sigma_1,\sigma_2, \ldots , \sigma_m)$ that are respectively 
mean and  variance of the output probability distribution from a given hidden
state (see Sections~\ref{sec:hmm} and \ref{sec:speecstudies}).  In our experiments we took the
default HTK values during the training step (i.e. $m = 25$ and $n =
5$).  To encode a single HMM we chose to focus only on the output
distributions, that is, the couple of vectors $(M,V)$. The idea 
is that all these values are initialized in the early steps of the training, according 
to a mean computed over the entire MFCC dataset: since all the values are iteratively 
refined through the HTK toolkit, then we expect that these values are correlated in some way
with the voices of the learning set and, by extension, with the
pronunciation dialects.
For this reason we set the feature vector $\vec{a} \in \mathcal{F}_{\mathcal{C}}$ as
follows:
$$
\vec{a} = (ph,\mu_1,\mu_2, \ldots , \mu_m,\sigma_1,\sigma_2, \ldots ,
\sigma_m,l_i)
$$ 
where $ph$ is a string value representing a phoneme, $\mu_1,\mu_2,
\ldots , \mu_m$ and $\sigma_1,\sigma_2, \ldots , \sigma_m$ are the
output probability vectors and $l_i \in \{$\emph{Indian},\emph{not Indian}$\}$ 
is the label of the current row. 
It is important to notice that this encoding gives a row in $\mathcal{D}_\mathcal{C}$ for 
each phoneme of the acoustic model.
Our training set was composed of $5,420$ tuples equally balanced over the 
two classifications considered for this experiment (i.e. the  50\% of training data
were generated by Indian people and the remaining 50\% by people speaking with 
different accent).
The test set was composed of $1,016$ instances: 774 of these are
classified as \emph{not Indian} and the remaining 242 are classified as
\emph{Indian}. The training ended up with a very complex meta-classifier:
the decision tree was composed of more than 811 nodes with 610 leaves.

\begin{table}
%\parbox{.45\linewidth}{
  \centering
  \begin{tabular}{ c | c || r }
    \hline                        
    Indian & not Indian & classified as \\
    \hline                        
    220  & 22 & Indian \\
    72 & 702 & not Indian\\

    \hline  
  \end{tabular}
  \caption{The confusion matrix of the meta-classifier\label{tab:conf-hmm1}}
\end{table}

%}
%\hfill
%\parbox{.45\linewidth}{
\begin{table}
  \centering
  \begin{tabular}{c || c | c}
  \hline
  & Precision & Recall \\
  \hline
  NotIndian & 0.97 & 0.91\\
  Indian &	0.75 & 0.91 \\
  \hline
  \end{tabular}
  \caption{The precision and recall summary  of the meta-classifier}
%}
\end{table}

Table \ref{tab:conf-hmm1} reports the confusion matrix obtained from
this experiment (we recall that the confusion matrix shows how
correctly a classifier assigned the labels to the elements of the
input set).  The \emph{not Indian} classifiers are correctly
classified with precision of 0.97 whereas
the \emph{Indian} classifiers are recognized with precision 0.75.
(Specifically: recall Indian: 0.909 and recall not Indian: 0.907.)\\
\begin{figure}
	\centering
	\includegraphics[width=0.5\textwidth,angle=-90]{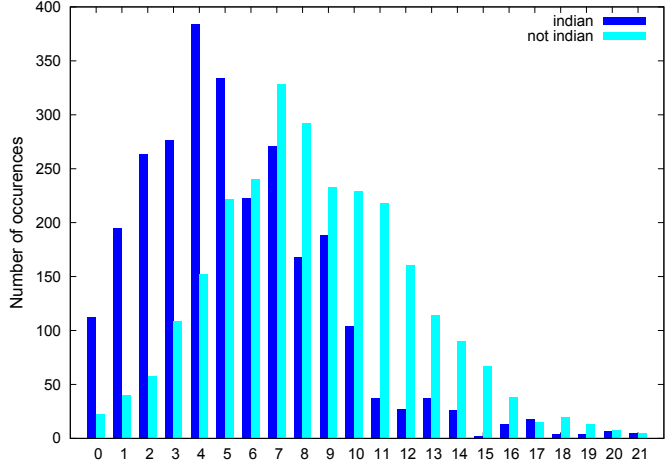}
	\caption{The frequency of the values of $\sigma_2$ for all phonemes in the training data of the meta-classifier.\label{fig:histogramsnotopt}}
\end{figure}
% TODO: rivedere l'inglese della frase sotto
One of the most interesting features provided by the \emph{C4.5} algorithm 
consists of the order in which the attributes decision tree appear. In fact
C4.5 puts the most representative attributes at the higher level of the tree.
In our experiment, one of the most representative nodes is $\sigma_2$.
The frequencies of each value of $\sigma_2$ in the training data of the meta-classifier 
are represented in figure~\ref{fig:histogramsnotopt}. It is easy to notice that the mean values 
of each distribution are considerably shifted and can be easily recognized with respect 
to the class.
Our meta-classifier is very effective in catching those differences; hence, as our experiments show, 
it correctly classifies  the most part of the test set. \\
To further improve the quality of $\mathbb{MC}$, we have applied a \emph{filter}
to the training set $\mathcal{D}_\mathcal{C}$. Our goal was to extract the 
phonemes that \emph{better differentiate} the language dialect.
To perform this task, we employed the \emph{Kullback-Leibler} (KL) divergence between the 
output probability distributions of the models.  
The KL divergence is defined as follows:
\begin{equation}
 D_{KL}(P||Q) = \sum_{i} P(i) log \dfrac{P(i)}{Q(i)}
\end{equation}
A low $D_{KL}$ value means a high similarity of the two probability
distributions, while on the other hand, high divergence values
correspond to an inferior similarity. This means that the phonemes with the
highest divergence are the ones which better discriminate the Indian
accent from others.\\ Since the output probabilities follow a Normal
distribution, we used the following equation to compute the KL divergence:
\begin{equation}
\mathfrak{D}_{KL}(X_i||X_j) = \frac{(\mu_i - \mu_j)^2 }{2\sigma^2_i} + \frac{1}{2}\left(\frac{\sigma^2_i}{\sigma^2_j}-1-\ln\frac{\sigma^2_i}{\sigma^2_j}\right)
\end{equation} 
where $X_i \sim N(\mu_i,\sigma_i)$ and $X_j \sim N(\mu_j,\sigma_j)$.\\
We built 100 different training sets without Indian records, obtaining
the relative acoustic models $\vec{\mathcal{C}} = (\mathcal{C}_1, \mathcal{C}_2,\ldots,\mathcal{C}_{100})$.
Then, we built the reference learning set containing only Indian records,
obtaining the relative acoustic model $\mathcal{C}_r$. Then, we compared
the distance between the output probability distributions of $\mathcal{C}_r$ with every 
$\mathcal{C}_i \in \vec{\mathcal{C}} $, obtaining the summed value of the
divergence. Since the same phoneme state has $25$ possible output distributions,
we have just computed the mean distance value across all the distributions.
Finally, we took the five phonemes with the highest divergence and we rebuilt
$\mathbb{MC}$ using only the entries relative to these phonemes.

\begin{table}
%\parbox{.45\linewidth} {
\centering
	\begin{tabular}{ c | c || r }
	  \hline                        
	    Indian & not Indian & classified as \\
	  \hline                        
	  169  & 6 & Indian \\	
	  2 & 137 & not Indian \\

	  \hline  
	\end{tabular}
	\caption{The confusion matrix of the \emph{filtered} meta-classifier\label{tab:conf-hmm2}}
\end{table}	

%	}
%	\hfill
%\parbox{.45\linewidth} {
\begin{table}

	\centering
	\begin{tabular}{c || c | c}
  \hline
  & Precision & Recall \\
  \hline
  NotIndian & 0.98 & 0.96\\
  Indian &	0.95 & 0.98 \\
  \hline
  \end{tabular}
  \caption{The precision and recall summary of the \emph{filtered} meta-classifier}
% }
\end{table}	
Table~\ref{tab:conf-hmm2} shows the confusion matrix of the
\emph{filtered} classifier. 
The new results are noticeably improved: the precision for the \emph{not Indian}
class is $0.98$ as before whereas the precision for
the \emph{Indian} class is increased to $0.95$. 
(Specifically: recall Indian: 0.986 and recall not Indian: 0.966.)\\
Also, the size of the 
decision tree has dropped down significantly (the resulting decision tree is composed only of
$21$ nodes with $11$ leaves).

\subsection{Support Vector Machines}
\label{sec:svm}
\subsubsection{Background}
Support Vector Machines (SVM) are supervised learning methods related
to \textit{statistical learning theory} and first introduced by Boser
et al. in~\cite{Boser92atraining}.\\
SVMs are largely used for classification and regression analysis. In
their basic form, SVMs are first trained with sets of input data
classified in two classes and are then used to guess the class for each
new given input.  This aspect makes SVM a
\textit{non-probabilistic binary linear classifier}.
Support Vector classifiers are based on the concept of \textit{separating hyperplanes}, 
that are the hyperplanes in the attribute space that defines the decision boundaries
between sets of objects belonging to different classes.

% \subsubsection{SVM Model}
% 
During the training phase, the SVM receives a set of labeled examples,
each of them described by $n$ numerical attributes (\textit{features})
and thus represented as a set of points in a $n$-dimensional space.
For the sake of simplicity, we briefly introduce how an SVM works with
data represented by two attributes and mapped into two classes.  The entry
$i$ of the training dataset is represented by a 2-dimensional vector
$x_i=\langle x_{i1}, x_{i2}\rangle$ and belongs to one and only one
class $y_i$:
\begin{equation}
\begin{split}
	(y_1,x_1), (y_2,x_2), \ldots  (y_n,x_m) \\
	y_j \in {-1,1}
\end{split}
\end{equation}
Let us suppose that the training data is linearly separable, namely
there exists a vector $w$ and a scalar value $b$ such that:
\begin{equation}
	\begin{split}
	\label{math:svm:inequality}
		w \cdot x_i\ +\ b\ \geq\ 1\ \textit{if}\ y_i\ =\ 1, \\
		w \cdot x_i\ +\ b\ \leq\ 1\ \textit{if}\ y_i\ =\ -1 
	\end{split}
\end{equation}
% Please notice that inequalities in~\ref{math:svm:inequality} can be 
% rewritten in the form:
% \begin{equation}
%  	y_i(w \cdot x_i\ +\ b)\ \geq\ 1
% \end{equation}
In order to deal with sets that are not linearly separable, the training vectors $x_i$ can be 
mapped
into a higher dimensional space by the function $\phi$, the so called
\textit{kernel function}: many kernel functions have been
proposed, but the most used  are linear $K(x_i,x_j) =
x^T_i x_j$, polynomial $K(x_i,x_j) = (\gamma x^T_i x_j + r )^d, \gamma
\ge 0 $, radial basis function, RBF, $K(x_i,x_j) = exp(-\gamma \|x_i
-x_j \|^2), \gamma \ge 0$ and sigmoid $K(x_i,x_j) = tanh(\gamma x^T_i
x_j + r)$.  The Support Vector classifier finds the optimal hyperplanes
that separate the training data with a maximal \textit{margin} in
this higher dimensional space; formally it resolves the
system of equations:
\begin{equation}
	\label{equation:margin}
 	y_i(w_0 \cdot x\ +\ b_0) = 0 
\end{equation}
It must be pointed out that, thanks to the nature of the training algorithm adopted by SVM,
the solution of (\ref{equation:margin}) can be obtained at a reasonable computational cost
regardless of the kernel function adopted.
% 
% Although new kernel funcion has been proposed by researcher, the most used and
% firstly presented mapping function are:
% \begin{itemize}
% 	\item linear: $K(x_i,x_j) = x^T_i x_j$
% 	\item polynomial: $K(x_i,x_j) =  (\gamma x^T_i x_j + r )^d, \gamma \ge 0 $
% 	\item radial basis function (RBF): $K(x_i,x_j) = exp(-\gamma \|x_i -x_j \|^2), \gamma \ge 0$
% 	\item sigmoid: $K(x_i,x_j) = tanh(\gamma x^T_i x_j + r)$
% \end{itemize}
% 
%  
%NOTA_AV: commentata perche non viene referenziata
%\begin{figure}[h]
%	\centering
%	\includegraphics[width=0.5\textwidth]{images/svm_intro}
%	\caption{Support Vector Classifier graphical interpretation:
%functional margin for a 2 class separable dataset.\label{fig:svm}}
%\end{figure}
%%
Intuitively, a good separation is achieved by the hyperplane that has
the largest distance - or {\it margin} - between the nearest training data points of different classes:  these points are called the {\it support vectors}.
Roughly speaking, the larger the margin, the lower  the generalization
error of the classifier.\\
It is easy to notice how the functional margin points determine 
the hyperplane of separation. This information is trivially 
featured by the attribute values in the training sets. Furthermore,
we highlight that SVM can disclose more information when several 
classifiers trained with different kernel functions are provided.
Since a trained SVM is represented by a set of weights and a subset
of the training sample, it is not easy to obtain useful information on
the characteristics of the complete training set directly from the SVM
representation.

% 
% USO DELL'SVM
% 
SVMs generated a significant research activity which extends across the limits of data
mining area. Although SVMs were initially introduced to solve pattern
recognition problems in an efficient way
(\cite{svm4patternrecognition}), nowadays they are suitable in several
contexts. In fact,  SVMs are used
for intrusion detection and anomaly detection 
(\cite{svm4IDS,SVM4AnomalyDetection,svm4nids}) or as part of complex
systems for similar tasks (\cite{svm4IDSandDT,svm4IDSandNN}). Other authors
propose SVM-based systems for privacy-critical tasks, such as cancer
diagnostic~\cite{svm4cancer,svmmedimage}, text categorization~\cite{svm4text}, or face recognition~\cite{svmface}.
%%%%%%
%%%%%
%%%%%
\subsubsection{SVM for network traffic classification}
\label{sec:casestudysvm}
As shown by the extensive literature on this topic
\cite{svm-tcp,Erman-semi,MLTrafficClassification,nettraffineural},
network traffic classification is commonly realized by means of
Machine Learning algorithms, like K-Means, HMM, decision trees, and SVM.  \\ In
order to evaluate the information leakage of SVM classifiers, we set
up a simple Network Traffic Classifier able to distinguish between DNS
and WEB traffic. In particular, we considered an SVM classifier based
on the \textit{SMO} module (Sequential Minimal
Optimization~\cite{SMO}) of the Weka framework.

% DESCRIZIONE DELL'AMBIENTE DELL'ESPERIMENTO, 
% LIBERAMENTE RIPRESO DALL'ARTICOLO REALDOS
Our experiment uses a \textit{real netflow} dataset,
gathered by a national tier 2 Autonomous System. 
%% TO UNCOMMENT WHEN NON-BLIND SUBMISSION:
% within the context of the Extrabire Project
% (\footnote{http://extrabire.eu}). This Autonomous System is
% connected to the three main network infrastructures present in Italy
% (Commercial, Research and Public Administration networks), and to
% several international providers.
NetFlow is a Cisco\texttrademark\ protocol used by network
administrators for gathering traffic statistics \cite{netflowrfc}.
NetFlow is used to monitor data at Layers 2-4 of the networking
protocol stack and to provide an aggregated view of the network
status. In particular, NetFlow efficiently supports many network tasks
such as traffic accounting, network billing and planning, as well as
Denial of Service monitoring.\\
A netflow-enabled router produces one new record for each newly
established connection, collecting selected fields from its IP header.
%(Figure~\ref{fig:netflow}). 
More precisely, a single netflow record
is defined as a \textit{unidirectional} sequence of packets all
sharing the following values: source and destination IP addresses, source
and destination ports (for UDP or TCP, 0 for other protocols), IP
protocol, Ingress interface index and IP Type of Service.\\
Other valuable information associated with the flow, such as timestamp,
duration, number of packets and transmitted bytes, are also recorded.
Then, we  consider a single netflow as a record that represents the
data exchanged between two hosts only in one direction.
We consider a network traffic classifier aimed at correctly
distinguishing the WEB and DNS traffic. 
The classifier was trained using a balanced set of netflows of WEB and DNS traffic.
It is worth noting that the WEB data set includes several traffic
patterns. Namely, it contains the flows directed to national
newspapers, advertising websites, and the Google search
engine website.\\
During the training phase of the experiment, we used all the fields of 
the netflow entries, except the source and destination IP addresses of 
the tracked connections.
In the literature there are examples of SVM
Classifiers for traffic detection~\cite{svm-tcp} able to 
distinguish a greater variety of network protocols; the methodology used 
in our experiment is similar, and can be considered appropriate to
highlight the statistical information leakage issues 
that are the target of our research.
Notice that the \textit{accuracy} and the \textit{precision} of the 
obtained classifier is optimal, thanks to the simplicity  
% artlessness 
of the training samples: indeed, WEB and DNS connections 
have well-separated traffic patterns, producing a large %AV 03-2013
margin for classification.
%producing a well-spaced support vector.

% SCOPO DELL'ESPERIMENTO
% The intuition of 
\subsubsection{Attack description}
In our experiment we investigate whether it is possible 
to extrapolate  the type of  traffic that was used during the
construction of the SVM model. For example: \textit{Can 
we infer whether Google web traffic was used in the training samples?}
(As before, Google traffic does NOT appear in the attributes of the 
training set.) 
We proceed with our attack by creating several ad-hoc data sets with well-defined statistical 
properties and use them to build our meta-classifier $\mathbb{MC}$.
% The sundry statistical features of the traffic patterns must be searched 
% in the nature of the connections: the network architecture, the traffic's 
% latency and other technical aspects contribute to characterize the flows. 
% These properties of the data set does not directly appears in the attributes 
% we consider during the learning phase, even if it are absorbed by the 
% classifiers. The amazing aspect  %[A]: forse questa frase è un po' troppo "spinta" ;-)
% of the attack is that we do not care about the statistical 
% properties of training sample of the traffic classifier we are 
% attacking, we train the Meta Classifier $\mathbb{MC}$ to learn it! 
% 
Namely, we created 70 ad-hoc data sets, selecting
20.000 flows of network traffic, distinct from the original training
set.
% directed to the website cited before.
% the selected entries 
% of the dataset were gathered by the same netflow collection by differ 
% from the data used during the building of $D_{all}$.
% In order to obtain traffic classifiers comparable with the target one, we 
% enrich each data sets with tantamount DNS traffic entries. 
While all 70 classifiers were trained with a non-specific DNS traffic,
the first half of the classifiers were trained using WEB traffic
directed only to Google search engine (property $\mathbb{P}$). For the
remaining 35 classifiers, we used WEB traffic without any netflow
directed to Google search engine (property $\overline{\mathbb{P}}$).

% The 70 classifiers builded reflects the property $\mathbb{P}$, \textit{the Google 
% web traffic has been used in the training samples} or the property $\overline{\mathbb{P}}$,
% \textit{the Google web traffic has not been used in the training samples}.
Each classifier was trained using a polynomial kernel function of
degree 3 and was encoded by the list of the support vectors it
contains, namely a set of points ($y,\vec{x}$) in the $n-$dimensional
space ($\vec{x} = \{x_1,x_2,\ldots,x_n\}$).  The training samples of
the classifier $\mathbb{MC}$ are composed of all the support vectors of
the 70 classifiers, labeled according to the property $\mathbb{P}$ or
$\overline{\mathbb{P}}$ used for training:
$$
\mathcal{D}_{\mathcal{C}} = \bigcup_{\mathcal{C}_{i}} \left\{
  (y,\langle x\rangle, label) \right\}
$$
We evaluate the performance of $\mathbb{MC}$ using the cross
validation strategy, a method that divides the data into $k$ mutually
exclusive subsets (namely, the ``folds'') of approximately equal
size. With cross validation, the accuracy estimate is the average
accuracy for the $k$ folds.
\begin{table}
%\parbox{.45\linewidth} {
\centering
	\begin{tabular}{ c | c | r }
	  \hline                        
	    Google & not Google & classified as\\
	  \hline                        
		2312  & 101 & Google\\
		92 	& 2786 & not Google\\
	  \hline  
	\end{tabular}
	\caption{The confusion matrix of the meta-classifier\label{tab:svmattack}}
\end{table}
%}
%\hfill
%\parbox{.45\linewidth} {
\begin{table}
\centering	
	\begin{tabular}{c || c | c}
  \hline
  & Precision & Recall \\
  \hline
  Google & 0.95 & 0.93\\
  not Google &	0.94 & 0.96 \\
  \hline
  \end{tabular}
  \caption{The precision and recall summary of the meta-classifier}
%}  
\end{table}

% NOTA: forse aggiungere considerazioni su quanto siano NASCOSTE queste caratteristiche?
Table~\ref{tab:svmattack} summarizes the experiment results: with
respect to the \textit{Google} class, we achieve a \textit{precision} 
of 0.954 and a \textit{recall} of 0.932. On the other hand, we correctly classify \textit{not Google}
instances with a \textit{precision} of 0.943 and a \textit{recall} of
0.962.

As in the example with the HMMs, the experimental results show that we
were able to build an effective meta-classifier that 
infers whether the training set given as input includes also a
specific type of traffic.

\section{Differential privacy}
\label{sec:differentialprivacyhacked}
In this section we show that differential privacy 
is ineffective against our attack strategy. More specifically, the information
leakage we are after sits outside the adversary model considered by differential privacy.

Differential privacy~\cite{dworkdiff,agrawal,sulq} protects
against unintentional disclosure of potentially sensitive information
related to a single record of a database $D$.
In other words, differential privacy maximizes the accuracy of queries
from statistical databases and, at the same time, minimizes the ability
to identifying single records.
To protect the privacy of database records, differential privacy opts for basically three
approaches:

\begin{enumerate}

\item The first is to obfuscate the original database $D$
and transform it into $D'$. This strategy is completely
ineffective in our model since $D'$ is the database actually used
during training and it is exactly what the adversary in our model is after.  
That is, our adversary is not interested in $D$, or any of its records, but it 
is rather eager for any information on $D'$, i.e., anything that is the result of
the transformations applied by differential privacy. 

\item Another approach is to train a classifier and
then add noise to the output. This is also ineffective since, in our model,
the adversary has complete access to the classifier and could just disable the
instruction that adds noise.

\item The third approach is more subtle. It consists 
of adding noise during training, thus effectively
obfuscating the learning process. This approach is still ineffective 
against our adversary since, intuitively, the final classifier
must anyway converge to classify correctly the training set.
Thus, the noise must be somehow restrained and its effect can easily 
be mitigated (see below). 
%Indeed, it can be shown that the added noise will be smaller than a
%given bound~\cite{sulq}, hence it is straightforward to mitigate its
%effect.
\end{enumerate}

It may be unclear why the third approach above fails to provide any protection against our adversary. Hence, we performed next an 
experiment showing how to extract sensitive information from a classifier trained 
within the framework SuLQ, introduced
in~\cite{sulq}.  The SulQ authors improved several standard classifiers 
to provide differential privacy. 
The main idea consists of adding a small amount of noise,
according to a Normal Distribution $N(0,\sigma)$, to any access to
the training set.
The variance of $N$ regulates the privacy property provided by differential privacy. 

Before introducing the experiment, we briefly recall some concepts of
K-Means, which is the most popular clustering algorithm.

\subsection{K-Means: the clusterization algorithm}
\label{sec:clustering}
% \subsection{A clusterization algorithm: k-means}

Clustering is the task of partitioning unstructured data in such a way
that objects with an high level of similarity fall into the same
partition.  Clustering is a typical example of unsupervised learning
models where examples are unlabeled, i.e., they are not
pre-classified.  The \textit{K-Means} algorithm~\cite{kmeans} is one
of the most common methods in this family and it has been used in many
applications (e.g.,~\cite{Maiolini-ssh,kmeansinfocom,Erman-kmeans,Salamatian-kmean}).
For example, in~\cite{Maiolini-ssh} the authors
developed a real-time traffic classification method, based on
\emph{K-Means}, to identify SSH flows from statistical behavior of IP
traffic parameters, such as length, arrival times and direction of
packets. 

In \emph{K-Means} both training and classification phases are very
intuitive. During the learning process, the algorithm partitions a set
of $n$ observations into $k$ clusters. Then, the algorithm selects the
\emph{centroid} (i.e., the barycenter, or geometric midpoint) of every cluster as a
representative for that set of objects.  More formally, given a set of
observations ($x_1, x_2, \ldots, x_n$), where each observation is a
$d$-\textit{dimensional}  real vector, \emph{K-Means} partitions the $n$
observations into $k$ sets $(k\le n)\ S = \{S_1, S_2, \ldots, S_k\}$
in order to minimize the within-cluster function:
\begin{equation}
  \underset{S}{\operatorname{argmin}} \sum_{i=1}^{k} \sum_{x_j \in S_i}^{} \| x_j - \mu_i \| \label{eq:kmeans}
\end{equation}
where $\mu_i$ is the mean of points in $S_i$.

To classify a given data set of $d$-\textit{dimensional} elements with
respect to $k$ clusters, \emph{K-Means} runs a learning 
%
%starts randomly choosing $k$
%centroids. Then, each point of the given data set is assigned to the
%cluster of the nearest centroid. With this assignment, $k$ new
%centroids are obtained evaluating the barycenters of the resulting
%clusters. If the $k$ new centroids induce a new assignment of the data
%points, $k$ new centroids are generated, and so on. The procedure is
%iterated until no modification in the assignment of data point to
%centroids is obtained from one iteration to the next.  The learning
process that can be summarized by the following steps:
\begin{enumerate}
\item Randomly pick $k$ initial cluster centroids; %: $c_1,$ $c_2,$ $c_3,$ $\ldots,$ $c_k$
\item Assign each instance $x$ to the cluster that has a centroid
  nearest to $x$;
\item Recompute each cluster's centroid based on which elements are
  contained in it;
\item Repeat Steps 2 and 3 until convergence is achieved;
\end{enumerate}
%% Angelo: QUESTA MI SEMBRA SBAGLIATA POICHÉ NON FA PARTE DI KMEANS
% During the classification phase, when a new observation $o$ must be
% classified, then the algorithm assign $o$ to the cluster with the
% nearest centroid.

\subsection{Hacking models secured by Differential Privacy}

We implemented two variants of a network traffic classifier that 
makes use of \emph{K-Means}. We
trained both classifiers with the same data set of the SVM experiment
of Section~\ref{sec:casestudysvm}.  The first implementation directly
uses the euclidian distance as metric to revise the \emph{centroids}
in the iterative refinement phase (equation~\ref{eq:kmeans}).  The
second version implements a privacy preserving version of
\emph{K-Means}, providing differential privacy. We implemented
the latter within  the SulQ framework, introduced by Blum et al.~\cite{sulq}.

We ran the two classifiers on $70$ training sets,  obtaining $70$
distinct centroids.  Recall that our objective is to recognize whether 
there was Google traffic within the traces. 

\begin{figure}
  \centering 
  \subfigure[Training set
  contains Web traffic directed to Google.com\label{fig:dp1}]{
    \label{fig:dp1}
	\includegraphics[width=0.7\textwidth]{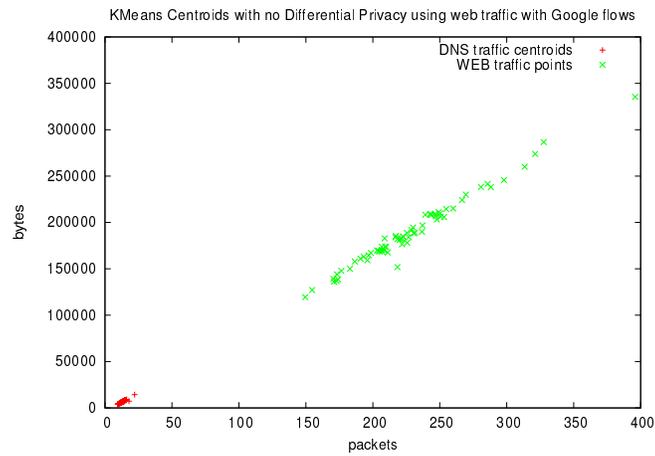}
      }
      \subfigure[Training set does not
      contain Web traffic directed to Google.com\label{fig:dp2}]{
        \label{fig:dp2}
	\includegraphics[width=0.7\textwidth]{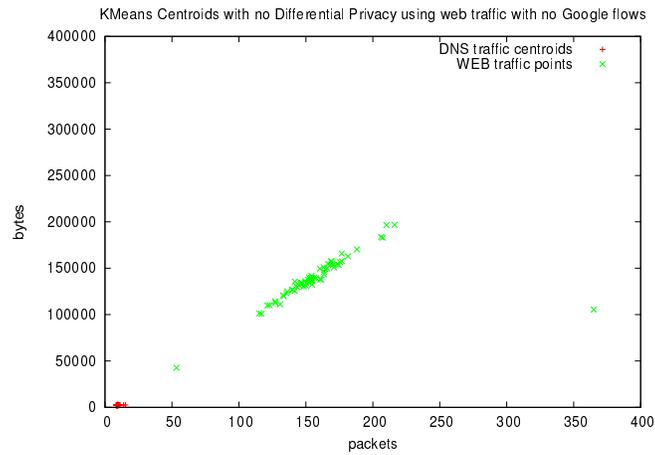}
      }
      \caption{Centroids of the
        K-Means traffic classifier {\em without} differential privacy.}
\end{figure}

With respect to the classifier with no differential privacy, we represent 
the centroids when there is traffic to Google.com in figure~\ref{fig:dp1},
and no traffic to Google.com in figure~\ref{fig:dp2}. 
It is easy to see that the positions of the centroids are quite different, allowing us 
to easily distinguish between these two cases. 

\begin{figure}
  \centering
  \subfigure[Training set contains Web traffic directed to 
Google.com\label{fig:dp3}]{
    \label{fig:dp3}
	\includegraphics[width=0.7\textwidth]{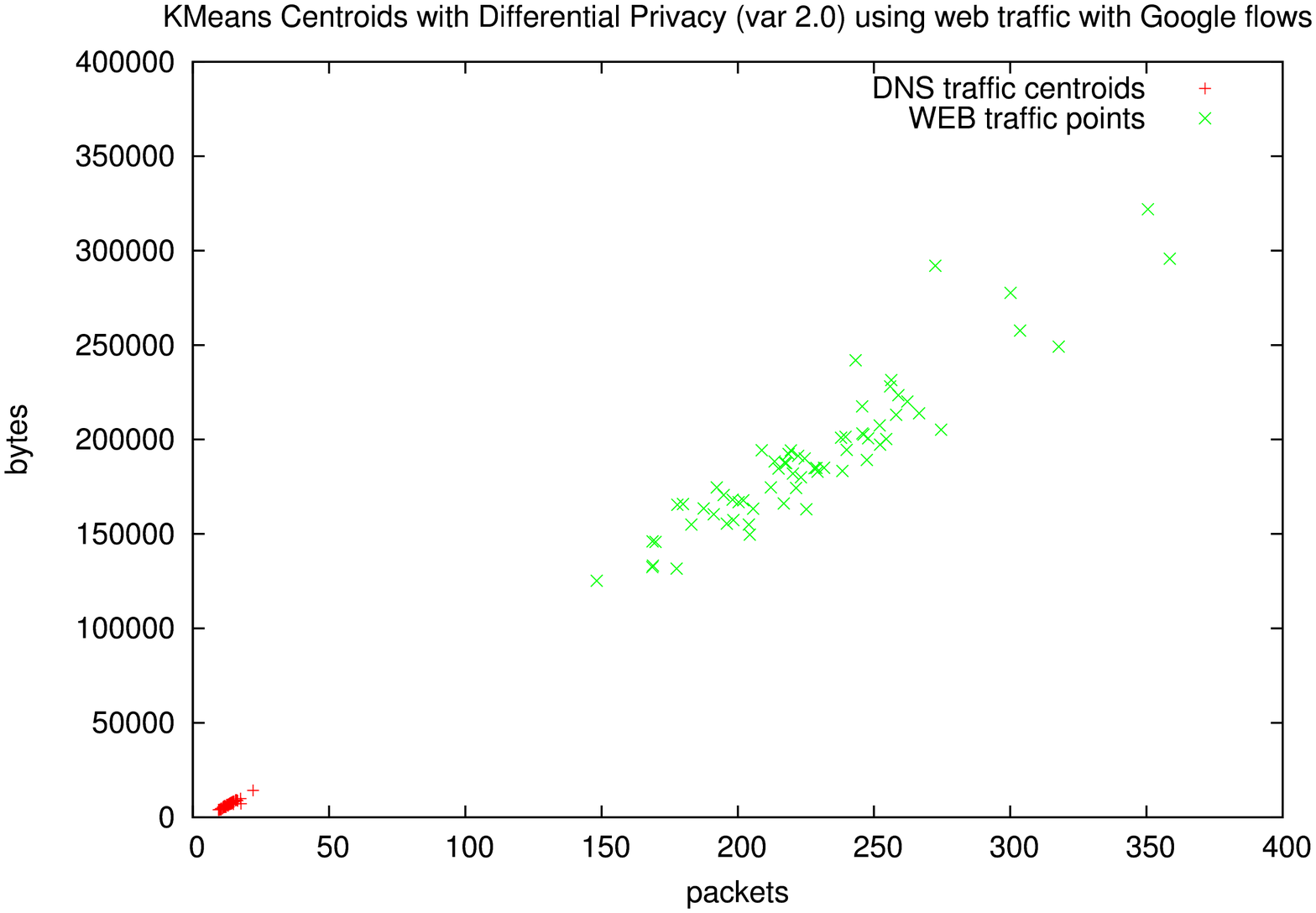}
  }
  \subfigure[Training set does not contain Web traffic directed to 
Google.com\label{fig:dp4}]{
    \label{fig:dp4}
	\includegraphics[width=0.7\textwidth]{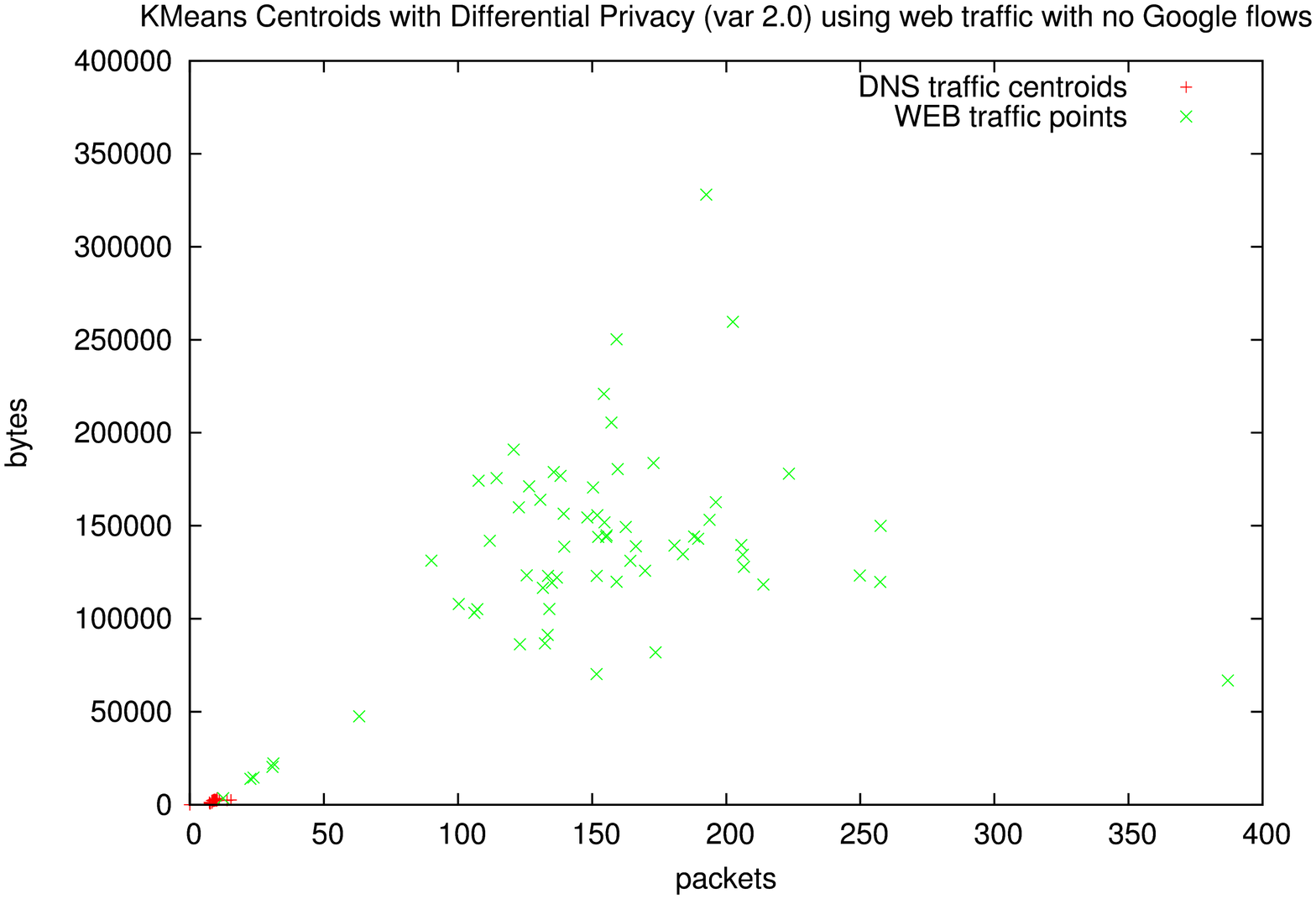}
  }
	\caption{Centroids of a K-Means Traffic Classifier {\em with} differential privacy.}
\end{figure}

Similar results appear when we picture the centroids of the classifier providing
differential privacy in figures~\ref{fig:dp3} and~\ref{fig:dp4}, respectively. 
Even in this case, an adversary can easily distinguish whether there is Google.com
traffic or not. 

% 
% SECTION RELATED WORKS 
% 
\section{Related works}
\label{sec:related}
The research area closest to the issues addressed in our paper appears
to be Information Disclosure considered in privacy preserving data
mining and statistical databases. It is worth describing
some of these related results, even though we stress that the type
of leakage we consider in this paper has not been considered before. 

% 
% DIFFERENTIAL PRIVACY
% 

As formalized by Dwork in~\cite{dworkdiff},  \textit{differential 
privacy} deals with the general problem of privacy preserving analysis 
of data.
More formally, \textit{a randomized mechanism $M$ provides $\epsilon-$differential
privacy if, for a database $D_1$ and $D_2$, which differ by at 
most one element, and for any t:}
$$
\frac{Pr[M(D_1)=t]}{Pr[M(D_2)=t]} \leq e^{\epsilon}
$$

In the differential privacy model, a trusted server holds a
database with sensitive information. Answers to queries
are perturbed by
the addition of random noise generated according to a random
distribution (usually a \textit{Laplace} distribution).\\
Two settings are defined: \textit{non interactive}, where 
the trusted server computes and publishes
statistics on the original data, and \textit{interactive}, where the server 
sits in the middle and directly alters the answers 
to user queries to guarantee specific privacy properties.\\
Chaudhuri et al. \cite{chaudhuri} design a privacy preserving logistic
regression algorithm which works in the $\epsilon-$differential
privacy model (\cite{calibratingnoise}).  The idea is quite simple: the
result of the trained classifier is perturbed with a dynamic amount of 
noise. This approach does not consider the security issues due to
the exposure of the model generated during the learning phase of the 
linear regression algorithm.\\
Other machine learning algorithms, such as Decision Trees, Artificial
Neural Networks, Clustering, have been re-engineered to provide 
differential privacy and several are defined within the
SulQ framework~\cite{sulq}.

%\textbf{TODO: arricchire questo inizio discorso (Giovanni)}
% Both Differential Privacy and privacy-preserving data mining, 
% survey several aspect of the privacy issues of statistical data
% and dataset.

%[AV] Toglierei la suddivisione in paragrafi che appesantisce un po' la lettura
% inserendo la frase seguente di colleggamento
\emph{Privacy Preserving Data Mining} (PPDM)~\cite{agrawal} is a novel research
area aimed at developing techniques that perform data mining
primitives while protecting the privacy of individual data records.
In~\cite{verykios}, Verykios et al. classified PPDM techniques in five classes.
Among them, we mention the \textit{Privacy preservation} class which 
refers to techniques used to preserve privacy for \textit{selective} modifications of data
records. 
This can be achieved through heuristic values (e.g., selecting the values that minimize the utility loss of the
  data), cryptographic protocols (e.g., via  Secure Multiparty
  Computation~\cite{smvprivpres}), or reconstruction-based techniques
  (e.g., strategy aimed at reconstructing  the original data distribution 
  using randomized data).\\
% extracting information about learning machines
%\begin{color}{red}
Some previous work exists related to extraction of information from classifiers.  
For instance, in~\cite{howtobeat04}, the authors show
how a bayesian learning algorithm can be used to learn which words are employed by a classifier 
to classify messages as spam and ham. Similarly, in~\cite{goodword05,Wittel04onattacking}, 
the authors describe some statistical attacks against spam filters aimed at understanding 
message features that are not correctly classified by the filters. 

Although using learning algorithms against other learning algorithms is not unprecedented(\cite{howtobeat04,aversarial-learning}), our approach is different since we 
uncovered a new class of information leakage that is inherent to the learning process and that has never been discussed before. 
% 
% SECTION CONCLUSION
% 
\section{Conclusions}\label{sec:conclusions}
In this paper we introduced a novel approach to extract meaningful data
from machine learning classifiers using a meta-classifier. 
While previous works investigated privacy concerns
of a single database record, our approach focuses on the statistical
information strictly correlated to the training samples used during the
learning phase.  We showed that several ML classifiers suffer from 
a new class of  information leakage that is not captured 
by privacy-preserving models, such as PPDM or
differential privacy. 

We devised a meta-classifier to successfully distinguish the accent
of users involved in defining the corpus of a
speech recognition engine.
Furthermore, we attacked an Internet
traffic classifier to infer whether a specific traffic pattern was
used during  training.  

Our results evince realistic  issues facing machine learning algorithms
as we put forward the importance of protecting the training set---the alluring 
recipe that makes a classifier better than the competition and that should 
be guarded as a trade secret.

\bibliographystyle{plain}
\bibliography{mlprivacy}

\newpage

\let\thefigureSAVED\thefigure  
\let\thetableSAVED\thetable

\appendix
\section{Artificial Neural Networks}
\label{sec:ann}
The \emph{Artificial Neural Networks} (ANNs) are a category of machine
learning algorithms able to solve a variety of problems in decision
making, optimization, prediction, and control, learning functions from
real, discrete and vector valued examples. 
The ANNs obtain good performances in problems where the training 
data is retrieved by complex sensor, such as cameras or microphones.
These algorithms are also resilient to the presence of
noise in the dataset.  Several types of ANN have been
proposed~\cite{Jain96artificialneural}. We focus on a particular family
 of ANNs, the
ones based on \emph{Multilayer Perceptrons}, and the related 
\emph{Backpropagation algorithm} (\cite{Chauvinbackprop}) 
used for their training.

The basic unit of an ANN is the \emph{Perceptron} (or neuron), a 
unit that takes a vector of real-valued inputs, calculates a linear
combination of these inputs and then outputs 1 if the result is
greater than some threshold and -1 otherwise. More formally a
perceptron can be represented as a function
\[
  o(x_1,\ldots,x_n) = \left\{ 
  \begin{array}{r l}
    1 & \quad \text{if $\sum_{i=0}^n w_ix_i > 0$}\\
   -1 & \quad \text{otherwise}\\
  \end{array} \right.
\]
where we consider $x_0$ to be always set to $1$ to simplify the
notation, and we call $\text{net} = \sum_{i=1}^n w_ix_i $. 
Observe that $-w_0$ is the threshold that makes the neuron to output 1. \\
A single perceptron represents an hyperplane decision surface in the
$n-dimensional$ space of instances. This kind of perceptron can only
discriminate between linearly separable instances. To overcome this
limitation, the sigmoid function $\sigma$ is used to decide the output
value:
$$    \sigma(net) = \frac{1}{1-\exp^{net}}  $$
An ANN is a multi-layer network of neurons: a first \textit{input}
layer receives the input bits and provides modified inputs to a
following layer, that, in turn, elaborates them and feeds a new layer,
and so on. The last layer outputs the result of the ANN. The neurons
that form the internal layers are called the hidden units.
The core function of the network resides in the weight of the hidden
units in the internal layers which are set through the \text{backpropagation} algorithm.
Starting
from random weights, the algorithm tunes them %the weights $w_0,\ldots,w_n$
using a training set of input-output pairs: the inputs go forward to
the network until they become output, while the errors (namely, the
difference between actual and expected outputs) are back-propagated 
to correct the weights. The error is reduced iteratively until a minimal and 
tolerable error is obtained. The backpropagation of
the error is inspired by the principle of gradient descent: in a nutshell,
if the weight  significantly contributes to the error then its adjustment will be
greater.

\section{Classification and Regression Trees}
\label{sec:tree}
A classification or regression tree (introduce by Breiman et al.
in~\cite{cl-rg-tree} in 1984) is a prediction model
which maps observations in a \textit{decision tree}.\\
The observations $L={(x_1,y_1), (x_2,y_2), \ldots, (x_N,y_N)}$
constitute the training set and are used to learn a decision tree.  
Both classification and regression
trees deal with the prediction of a response variable $y$ (let $Y$ be the
domain of $y$), given the values of a vector of predictor variables
$x$ (let $X$ be the domain of $x$).  If $y$ is a continuous or discrete
variable taking real values (e.g., the size of an object, the number
of occurrences of certain events), the problem is called
\textit{regression}; if $Y$ is a finite set of
unordered values (e.g., the type of Iris plants), the problem is
called \textit{classification}.  \\ The training phase produces a tree
structure in which the leaves represent the class labels and the
branches represent \textit{conjunctions of features} that lead to the
class labels of their leaves.  Decision trees can be considered as
disjunction of conjunctions of constraints on the attribute-values of
instances. Each path from the tree root to a leaf corresponds to a
conjunction of attribute tests, and the tree itself to a disjunction
of these conjunctions~\cite{mitchell-ml}.  Decision trees work better
when the target function has discrete output (for example ``yes or
no'') and the data instances are represented by attribute-value pairs.
Furthermore, decision trees perform well even when the training
dataset contains errors or missing values.  These characteristics make
decision tree a suitable solution for many classification problems 
and in a great variety of contexts.
The most popular implementation of  decision trees is the $C4.5$~\cite{QuinlanC45}
algorithm, which is an extended version of the $ID3$ algorithm~\cite{Quinlan:introDT}.
top-down, greedy search through the space of all possible decision
trees. In detail, \textit{ID3} algorithm starts the search of decision
tree answering the question: \textit{which attribute should be used at
  the root of the tree?} Once the root is found, a descendent node of
the root is created for each possible value, then the same question
is asked recursively at each new node, until: (i) each attribute has been
considered in the path through the tree, or (ii) the training examples
related to a specific leaf has the same attribute values.  The
selection of the best attribute in each level of the tree is performed
using the concept of \textit{information gain}. In fact, the
information gain measures how well a given attribute separates the
training examples. Given a collection $S$ of items, for each
attribute $A$, $ID3$ algorithm evaluates the gain of $A$ with respect
to $S$ via the equation:
$$
\textit{Gain}(S,A) = H(S) - \sum_{v \in {\small \textit{Values}}(A)} \dfrac{|S_v|}{|S|}
H(S_v)
$$
where $H(S)$ represents the Entropy of the entire dataset and $S_v$ is
the subset of $S$ for which attribute $A$ has value $v$.  
%Roughly
%speaking, this measure represents the expected reduction in entropy
%caused by partitioning the training set using the attribute under
%analysis.

\let\thefigure\thefigureSAVED 
\let\thetable\thetableSAVED

\end{document}